\documentclass[12pt]{article}
\usepackage{epsf}
\title{ {\bf
Dark matter as a localized scalar in the extra dimension}}
\author{\vspace{1cm}\\
        {\bf E. O. Iltan}
        \thanks{E-mail address:
        eiltan@newton.physics.metu.edu.tr}
\\
Physics Department, Middle East Technical University \\
        Ankara, Turkey \\
        }

\date{}

\begin{document}
\setlength{\baselineskip}{24pt}
\maketitle
\setlength{\baselineskip}{7mm}
\begin{abstract}
We consider a standard model singlet which is accessible to a
single extra dimension and its zero mode is localized with
Gaussian profile around a point different than the origin. This
zero mode scalar is a possible candidate for the dark matter and
its annihilation rate is sensitive to the compactification radius
of the extra dimension, the localization width and the position.
For the case of non resonant annihilation, we estimated the dark
matter scalar location around a point, at a distance $\sim 3\,
\times$ localization width from the origin, by using the
annihilation rate which is based on the current relic density.
\end{abstract}
\thispagestyle{empty}
\newpage
\setcounter{page}{1}
The amount of matter required in the universe is considerably
greater than the visible one and there is a need to construct a
theoretical background in order to explain the excess invisible
matter, called as dark matter (DM). The galactic rotation curves
\cite{BorrielloA} and galaxies orbital velocities \cite{ZwickyF},
the cosmic microwave background anisotrophy \cite{WAMP2},
observations of type Ia supernova  \cite{KomatsuE} indicate that
almost $23\%$ of present Universe \cite{KomatsuE, JungmanG2,
JungmanG} is composed of cold (non relativistic) DM, however, its
nature is not known. In the literature, there are various attempts
to solve the DM problem; the DM candidates in supersymmetric
models \cite{MAsano}, in the universal extra dimension (UED)
models \cite{Cheng1, Cheng2, Matsumoto}, split UED models
\cite{ServantG, Park, Nojiri}, Private Higgs model \cite{Jackson},
Inert doublet model \cite{MaE}-\cite{Calmet}, Little Higgs model
\cite{Bai}, Heavy Higgs model \cite{Majumdar}. Among the possible
candidates the Weakly Interacting Massive Particles (WIMPs) that
interact only through weak and gravitational interactions and have
masses in the range $10$ GeV- a few TeV reached great interest
since the current relic density could be explained by thermal
freeze-out of their pair annihilation (see for example
\cite{DEramoF, WanLeiGuo} for further discussion).

In the present work we consider a scalar SM singlet $\phi_S$ which
is accessible to a single extra dimension and all SM particles
live on the 4 dimensional brane\footnote{See \cite{SilveiraV} and
\cite{Burgess, HolzE, McDonaldJ, PattB, BertolamiO, DavoudiaslG,
HeXG} for an additional scalar SM singlet field $\phi_S$ in 4
dimensions, called as darkon}. After the compactification of the
extra dimension, denoted by $y$, on an orbifold $S^1$ with radius
$R$ the KK modes of the SM singlet scalar appear. At this stage,
we assume that the zero mode scalar is localized\footnote{We
consider that the mechanism for the localization is unknown.} (see
for example \cite{Eoi1} for a special localization mechanism for a
scalar in the  extra dimension) at the point $y_L$ in the extra
dimension, with the Gaussian profile as
\begin{eqnarray}
\phi^{(0)}_S=f_L(y)\, S(x) \label{phi2L}\, ,
\end{eqnarray}
where the function $f_L(y)$ reads
\begin{eqnarray}
f_L(y)=A_L \,e^{-\beta (y-y_L)^2} \label{fL}\, .
\end{eqnarray}
The normalization constant $A_L$ is obtained by using the integral
\begin{eqnarray}
\int_{-\pi\,R}^{\pi\,R}\,f_L^2(y)\,dy=1 \label{intnorm}\, ,
\end{eqnarray}
and one gets
\begin{eqnarray}
A_L=\frac{2\,( \beta)^{1/4}}{(2
\pi)^{1/4}\,\sqrt{Erf[\sqrt{2\,\beta}\,(\pi\,R+y_L)]+
Erf[\sqrt{2\,\beta}\,(\pi\,R-y_L)]}} \label{NormH} \, .
\end{eqnarray}
Here $\beta$, $\beta=1/\sigma^2$, is the parameter which adjusts
the localization amount of $\phi^{(0)}_S$ and $\sigma$ is the
Gaussian width of $\phi^{(0)}_S$ in the extra dimension. The
function $Erf[y]$ is the error function, which is defined as
\begin{eqnarray}
Erf[y]=\frac{2}{\sqrt{\pi}}\,\int_{0}^{y}\,e^{-t^2}\,dt \,\, .
\label{erffunc}
\end{eqnarray}
Furthermore, we assume that the lagrangian of  the scalar SM
singlet has the discrete $Z_2$ symmetry, $\phi^{(0)}_S\rightarrow
-\phi^{(0)}_S$ in order to ensure its stability and has the
interaction term in a single extra dimension\footnote{Here an ad
hoc non zero mass term for the zero mode singlet $\phi^{(0)}_S$ is
considered and it depends on the localization mechanism which is
unknown in our case. Notice that the electroweak symmetry breaking
also leads to a contribution to the tree level mass due to the
interaction term (see eq.(\ref{Vint})).}
\begin{eqnarray}
{\cal{L}}_{Int}=\lambda_{5\,S}\, \phi^{(0)\,
2}_S|_{y=0}\,(\Phi_1^\dagger\, \Phi_1)\, ,\label{Vint}
\end{eqnarray}
where $\Phi_1$ is the SM Higgs field
\begin{eqnarray}
\phi_{1}=\frac{1}{\sqrt{2}}\left[\left(\begin{array}{c c}
0\\v+H^{0}\end{array}\right)\; + \left(\begin{array}{c c} \sqrt{2}
\chi^{+}\\ i \chi^{0}\end{array}\right) \right]\, ,
\label{HiggsField}
\end{eqnarray}
with the vacuum expectation value
\begin{eqnarray}
<\phi_{1}>=\frac{1}{\sqrt{2}}\left(\begin{array}{c c}
0\\v\end{array}\right) \, . \label{VEV}
\end{eqnarray}
Here we consider that the SM singlet scalar $\phi_S$ has no vacuum
expectation value and, with the discrete $Z_2$ symmetry above, it
is guaranteed that the scalar field $S$ has no SM decay products.
After the electroweak symmetry breaking, they disappear by pair
annihilation with the help of the exchange particle which is the
SM Higgs boson $H^0$ in this case. Therefore, the zero mode scalar
is a possible candidate for DM and we will study the effects of
compactification radius, the zero mode scalar localization width
and its localization position on the annihilation cross section.

The annihilation process $S\,S\rightarrow H^0\,\rightarrow X_{SM}$
results in the total averaging annihilation rate of $S\,S$
\begin{eqnarray}
<\sigma\,v_r>&=&\frac{4\,\lambda_S^2\, v^2}{m_S}\,\frac{1}{
(4\,m_S^2-m_{H^0}^2)^2+m^2_{H^0}\,\Gamma^2_{H^0}}\,
\Gamma(\tilde{h}\rightarrow X_{SM})\, ,  \label{sigmavr}
\end{eqnarray}
where $\Gamma(\tilde{h}\rightarrow
X_{SM})=\sum_i\,\Gamma(\tilde{h}\rightarrow X_{i\,SM})$ with
virtual Higgs $\tilde{h}$ having mass $2\,m_S$ (see \cite{BirdC,
BirdC2}) and $v_r=\frac{2\,p_{CM}}{m_S}$ is the average relative
speed of two zero mode scalars (see for example \cite{HeXG}). In
eq.(\ref{sigmavr}) the effective coupling $\lambda_S$ reads
\begin{eqnarray}
\lambda_S=\lambda_{5\,S}\,A_L^2 \,e^{-2\,\beta \,y_L^2}\, .
\label{lambdaS}
\end{eqnarray}
Here, we parametrize  $\sigma$ and $y_L$ as $\sigma=\rho\,R$,
$y_L=\alpha\, \sigma$ where $\rho$ and $\alpha$ are dimensionless
parameters and, for the effective coupling $\lambda_S$, we get
\begin{eqnarray}
\lambda_S=\frac{4\,e^{-2\,\alpha^2}}{\sqrt{2\,\pi}\,\rho\,\,R\,
\Big( Erf[\sqrt{2}\,(\frac{\pi}{\rho}+\alpha)]+
Erf[\sqrt{2}\,(\frac{\pi}{\rho}-\alpha)]\Big)}\, \,\,,
\label{lambdaS2}
\end{eqnarray}
with the choice $\lambda_{5\,S}=1.0\,GeV^{-1}$. Therefore, in this
scenario, the strength of tree level interaction of the scalar DM
pair with the SM Higgs boson is regulated by the scalar DM
localization point, its localization width and the
compactification radius of the extra dimension. In order to
estimate these parameters one needs a restriction for  the
annihilation cross section $< \sigma\,v_r>$.  The present DM
abundance by the WMAP collaboration \cite{WAMP} is
\begin{eqnarray}
\Omega\,h^2=0.111\pm 0.018 \, , \label{RelDens}
\end{eqnarray}
and the annihilation cross section is inversely proportional to
the relic density as
\begin{eqnarray}
\Omega\,h^2=\frac{x_f\,10^{-11}\,GeV^{-2}}{< \sigma\,v_r>}
\,,\label{omegahsig}
\end{eqnarray}
where $x_f\sim 25$ \cite{JungmanG2, ServantG, HeXG, Gopalakrishna,
Gopalakrishna2}. Finally, eqs.(\ref{RelDens}) and
(\ref{omegahsig}) lead to the bounds for the annihilation cross
section,
\begin{eqnarray}
< \sigma\,v_r>=0.8\pm 0.1 \, pb \,,\nonumber
\end{eqnarray}
in the case that  s-wave annihilation is dominant (see
\cite{KolbEW} for details.).
%
\\ \\
{\Large \textbf{Discussion}}
\\ \\
%
In the present work we introduce a scalar field, which is a SM
singlet, with vanishing vacuum expectation value and consider that
its zero mode localized with Gaussian profile around a point away
form the origin, in the extra dimension. The interaction
Lagrangian (see eq. (\ref{Vint})) results in that this field
interacts with the SM Higgs $H^0$ and the tree level interaction
$S\,S\,H^0$ arises with the effective coupling $\lambda_S$
(eq.(\ref{lambdaS2})) after the elecroweak symmetry breaking. This
coupling drives the annihilation cross section which should be
compatible with the present observed DM relic density
(eq.(\ref{RelDens})). The free parameters of the model used are
the Higgs mass $m_{H^0}$, the zero mode scalar mass $m_S$, the
compactification radius $R$ of the extra dimension, the
localization position $y_L$ of the zero mode scalar and its
localization width $\sigma$. Here, we take Higgs mass around
$110-120\,GeV$ and use the range $10-80\, GeV$ for the zero mode
scalar mass. Furthermore, we respect the prediction of the present
DM abundance and use the central value of $< \sigma\,v_r>=0.8\,pb$
in order to predict the localization position of the DM with
respect to the compactification radius $R$ and the DM mass $m_S$.
In the numerical calculations we take the compactification radius
$R$ in the range $0.00001\,GeV^{-1} \leq \, R\,
\leq\,0.001\,GeV^{-1}$. In the range of free parameters we choose,
the coupling $\lambda_S$ obeys $\lambda_S< 1.0$, which is
necessary for perturbative calculations. Notice that the direct
detection experiments ensure an upper limit (see \cite{Akerib} for
the current upper limit) for the WIMP-nucleon cross
section\footnote{The spin independent cross section can be given
as
$\sigma=\frac{f^2\,m_{N}^2\,m_{SN}^2}{4\,\pi}\,(\frac{\lambda_S}{m_S\,
m^2_{H^0}})^2$ where $m_N$ ($m_{SN}$) is the nucleon mass (reduced
mass of $S$ and nucleon) and the coupling $f$ reads $f\sim 0.3$
(see for example \cite{Hall, APierce}).}. The parameter set we
used leads to the cross section for proton target at the order of
magnitude of $10^{-8}-10^{-7}\, pb$ which is almost one order less
than the current limit. This prediction shows that the considered
parameter set, which respects the bounds of the annihilation cross
section, is not rule out by the results of direct detection that
might be improved with the forthcoming experiments.

\begin{figure}[htb]
\vskip -3.0truein \centering \epsfxsize=6.8in
\leavevmode\epsffile{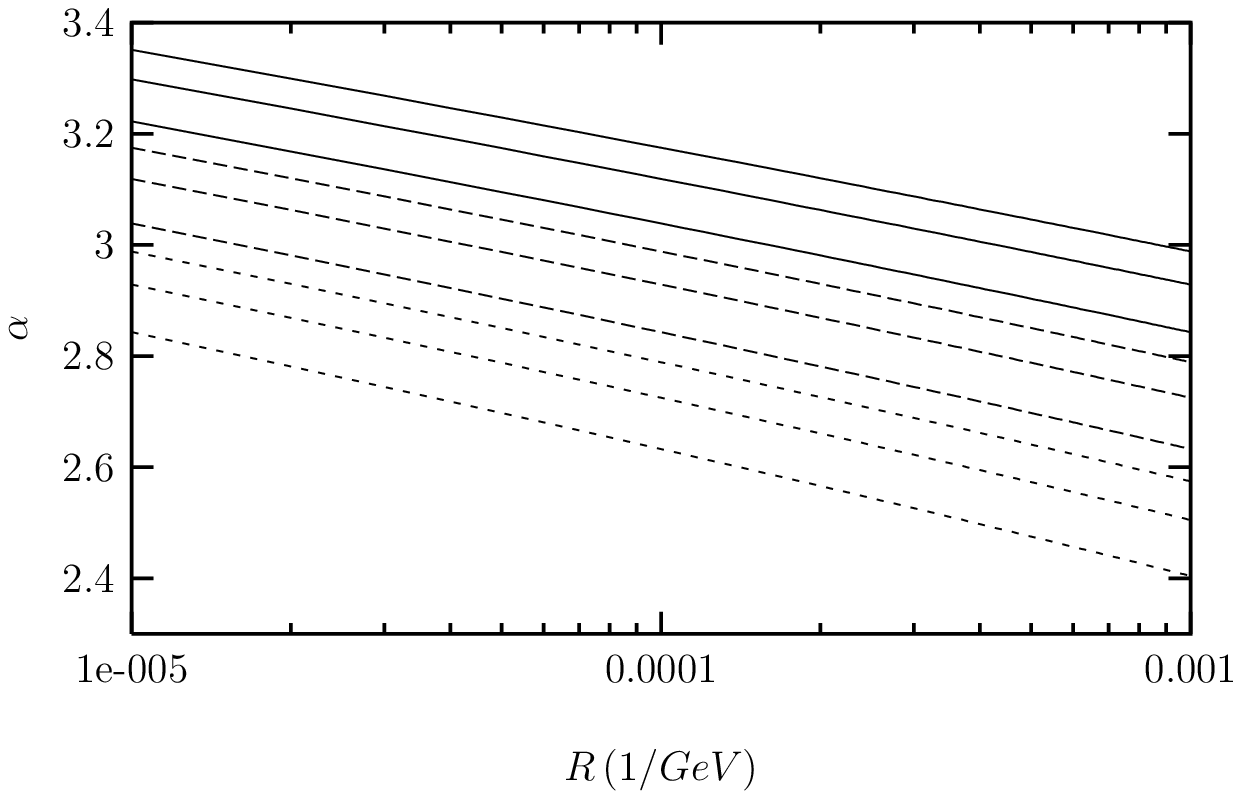} \vskip -3.0truein \caption[]{
$\alpha$ as a function of $R$ for $m_{H^0}=110\,GeV$. Here the
upper-intermediate-lower solid (long dashed; dashed) line
represents $\alpha$ for $m_S=50-80-10\,GeV$, $\rho=0.001\,(0.01;
0.1)$.} \label{alfaR}
\end{figure}

In Fig.\ref{alfaR} we plot the compactification radius $R$
dependence of $\alpha$ for $m_{H^0}=110\,GeV$. Here the
upper-intermediate-lower solid (long dashed; dashed) line
represents $\alpha$ for $m_S=50-80-10\,GeV$, $\rho=0.001\,(0.01;
0.1)$. It is observed that the parameter $\alpha$, which measures
the distance of the localization point of the DM scalar from the
origin in the extra dimension, is in the range of
$2.4\,\sigma-3.3\,\sigma$, for the chosen interval of the free
parameters. For $m_S=50\, GeV$, the mass of the scalar which is
near to the resonant annihilation, the interaction coupling
$\lambda_S$ should be regulated to a small value in order to reach
the appropriate annihilation cross section which agrees with the
current relic density. This is the case that the DM scalar is
localized far from the origin in order to weaken the interaction
with the Higgs particle and, therefore, the parameter $\alpha$
reaches to relatively greater values. In the case of the DM with
the mass value far from the resonant annihilation, the heavier DM
has a weak cross section compared to the lighter one and, in order
to satisfy the observed relic abundance, it should be pulled to
the appropriate value by choosing the weaker coupling, i.e.
relatively larger $\alpha$. This figure shows that $\alpha$ is
sensitive to the parameter $\rho$, which fixes the width of the
localization, and to the compactification radius $R$. With the
increasing values of $\rho$ and $R$ $\alpha$
decreases\footnote{Notice that the width of the localization which
is  parameterized by $\rho$ is chosen at least one order less than
the compactification radius $R$}.

\begin{figure}[htb]
\vskip -3.0truein \centering \epsfxsize=6.8in
\leavevmode\epsffile{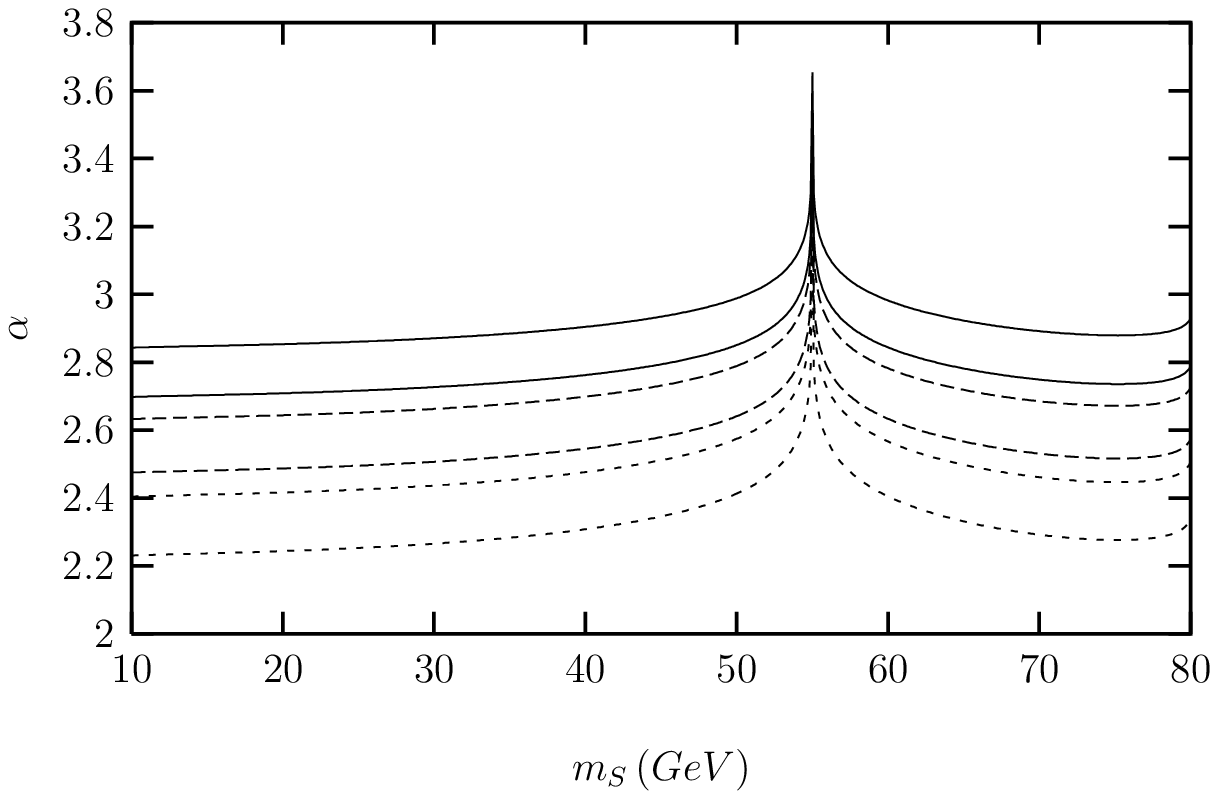} \vskip -3.0truein \caption[]{
$\alpha$ as a function of $m_S$ for $m_{H^0}=110\,GeV$. Here the
upper-lower solid (long dashed; dashed) line represents $\alpha$
for $R=0.001-0.005\,GeV^{-1}$, $\rho=0.001\,(0.01; 0.1)$.}
\label{alfamS}
\end{figure}

Fig.\ref{alfamS} represents the DM scalar mass $m_S$ dependence of
$\alpha$ for $m_{H^0}=110\,GeV$. Here the upper-lower solid (long
dashed; dashed) line represents $\alpha$ for
$R=0.001-0.005\,GeV^{-1}$, $\rho=0.001\,(0.01; 0.1)$. We observe
that the parameter $\alpha$ reaches to the largest value $\sim
3.6\,\sigma$ in the case of resonant annihilation and decreases
when the DM mass becomes far from the resonance mass,
$m_S=55\,GeV$. Near the resonant case $\alpha$ is sensitive the DM
scalar mass and this sensitivity becomes weak when the mass $m_S$
is far from $55\,GeV$, in the range of $m_S$ considered. It is
seen that, for the large mass values, $m_S\geq 80\,GeV$, this
sensitivity becomes strong and $\alpha$ increases, since the
annihilation cross section enhances and it should be suppressed by
appropriate weak coupling which is regulated by the parameter
$\alpha$.

\begin{figure}[htb]
\vskip -3.0truein \centering \epsfxsize=6.8in
\leavevmode\epsffile{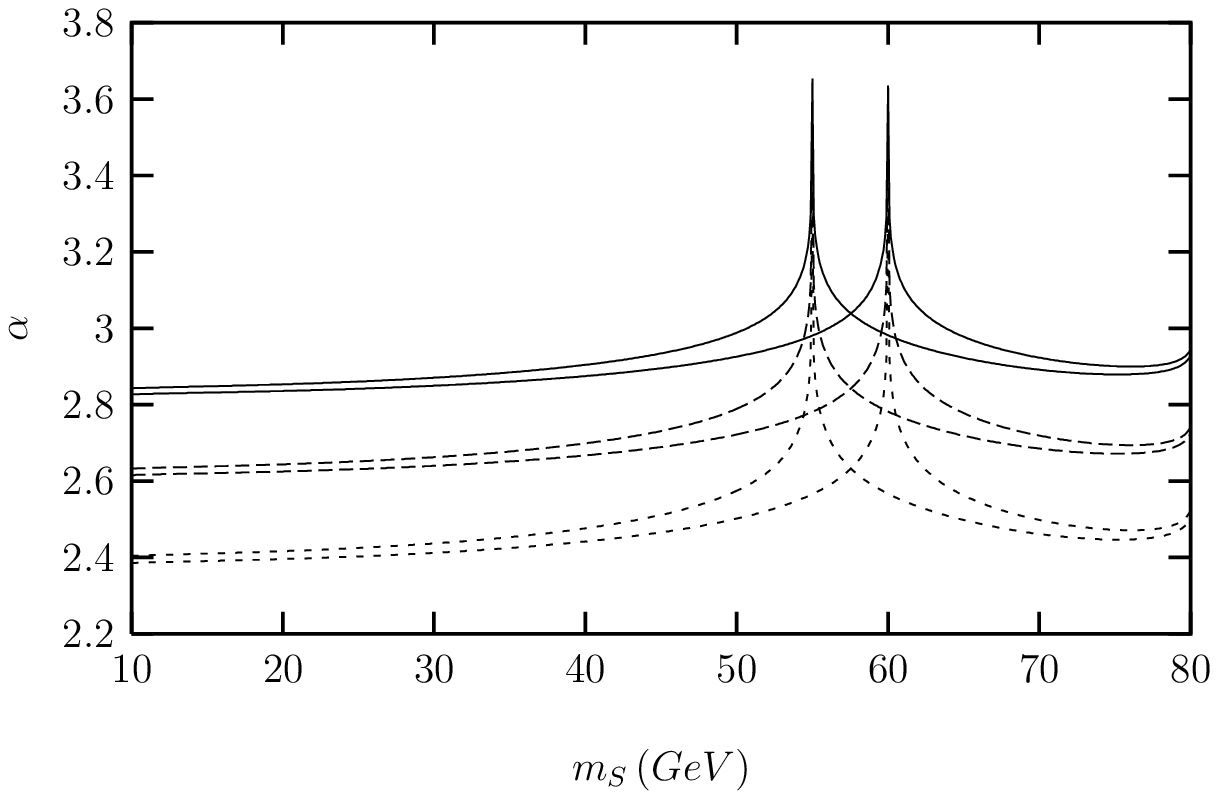} \vskip -3.0truein \caption[]{
$\alpha$ as a function of $m_S$ for $R=0.001\,GeV^{-1}$. Here the
left-right solid (long dashed; dashed) line represents $\alpha$
for $m_{H^0}=110-120\,GeV$, $\rho=0.001\,(0.01; 0.1)$.}
\label{alfamS2}
\end{figure}

In Fig.\ref{alfamS2} we present the DM scalar mass $m_S$
dependence of $\alpha$ for $R=0.001\,GeV^{-1}$. Here the
left-right solid (long dashed; dashed) line represents $\alpha$
for $m_{H^0}=110-120\,GeV$, $\rho=0.001\,(0.01; 0.1)$. Here two
maximum values of $\alpha$ at different $m_S$ are due to two
different resonant annihilations, namely the annihilations for
$m_S=55\,GeV$ and $m_S=60\,GeV$. This figure shows that $\alpha$
is greater (smaller) almost for $m_S> 57\,GeV$ ($m_S< 57\,GeV$) in
the case of $m_{H^0}=120\,GeV$ compared to the case of
$m_{H^0}=110\,GeV$. With the increasing values of the Higgs mass,
the localization point of the scalar DM in the extra dimension
goes far from (comes near to) the origin if the scalar DM mass
$m_S$ is greater (less) than the resonant annihilation mass.

As a summary, we consider that the additional scalar field is
accessible to a single extra dimension and its zero mode is
localized. With the ad hoc symmetry in the Lagrangian assumed this
zero mode becomes stable and annihilates to the SM Higgs particles
and, therefore, it is a candidate for a scalar DM. In this
scenario, we estimate the position of the localization point of
the scalar DM in the extra dimension by respecting the current
relic density. We observed that the localization point of the DM
scalar places at $2.4\,\sigma-3.4\,\sigma$ distance from the
origin in the extra dimension, for the chosen interval of the free
parameters. In the case of the resonant annihilation this distance
reaches to $3.6\,\sigma$. Furthermore we see that this distance
decreases with the increasing values of localization width
parameter $\rho$ and compactification radius $R$ and it is
sensitive to the SM Higgs boson mass. Hopefully, the observation
of the SM Higgs boson in the future experiments at LHC will
provide a considerable information about the nature of the DM and
the possible mechanism which drives the DM-SM Higgs annihilation.
\newpage
\newpage

\begin{thebibliography}{1}
%
\bibitem{BorrielloA} A. Borriello and P. Salucci,
{\it Mon. Not. Roy. Astron. Soc.} {\bf 323}, 285 (2001).
%
\bibitem{ZwickyF} F. Zwicky,
{\it Helv. Phys. Acta.} {\bf 6}, 110 (1993).
%
\bibitem{WAMP2} D. N. Spergel et.al (WAMP Collaboration),
{\it Astrophys. J. Suppl. Ser.} {\bf 170} 377 (2007).
%
\bibitem{KomatsuE} E. Komatsu et al., {\it Astrophys. J. Suppl. Ser.}
{\bf 180}, 330 (2009).
%
\bibitem{JungmanG2} G. Bertone, D. Hooper and J. Silk, {\it Phys. Rept.}
{\bf 405}, 279 (2005).
%
\bibitem{JungmanG} G. Jungman, M. Kamionkowski and
K. Griest, {\it Phys. Rept.} {\bf 267}, 195 (1996).
%
\bibitem{MAsano} M. Asano, S. Matsumoto, M. Senami, H. Sugiyama,
{\it Phys. Lett.} {\bf B663}, 330 (2008).
%
\bibitem{Cheng1} T. Appelquist, H. C. Cheng and B. A. Dobrescu,
{\it Phys. Rev.} {\bf D64}, 035002 (2001).
%
\bibitem{Cheng2} H. C. Cheng, K. T. Matchev and M. Schmaltz,
{\it Phys. Rev.} {\bf D66}, 056006 (2002).
%
\bibitem{Matsumoto} S. Matsumoto, J. Sato, M. Senami, M. Yamanaka,
hep-ph/0903.3255.
%
\bibitem{ServantG} G. Servant and T. M. P. Tait,
{\it Nucl. Phys.}  {\bf B650}, 391 (2003).
%
\bibitem{Park} S. C. Park and J. Shu, {\it Phys. Rev.} {\bf D79},
091702(R) (2009).
%
\bibitem{Nojiri} C. R. Chen, M. M. Nojiri, S. C. Park, J. Shu, M. Takeuchi,
hep-ph/0903.1971.
%
\bibitem{Jackson}  C.B. Jackson, hep-ph/0804.3792.
%
\bibitem{MaE} E. Ma, {\it Phys. Rev.} {\bf D73}, 077301 (2006).
%
\bibitem{Hall} R. Barbieri, L. J. Hall, and V. S. Rychkov,
{\it Phys. Rev.} {\bf D74}, 015007 (2006).
%
\bibitem{Cirelli} M. Cirelli, N. Fornengo, and A. Strumia,
{\it Nucl. Phys.} {\bf B753}, 178 (2006).
%
\bibitem{Deshpande}  N. G. Deshpande and E. Ma,
{\it Phys. Rev} {\bf D18}, 2574 (1978).
%
\bibitem{Casas} J. A. Casas, J. R. Espinosa, and I. Hidalgo,
{\it Nucl. Phys.} {\bf B777}, 226 (2007).
%
\bibitem{LHonorez} L. L. Honorez, E. Nezri, J. F. Oliver, M. H. G. Tytgat,
{\it JCAP} {\bf 0702}, 028 (2007).
%
\bibitem{SAndreas} S. Andreas, T. Hambye, M. H.G. Tytgat,
{\it JCAP} {\bf 0810}, 034 (2008).
%
\bibitem{Calmet} X. Calmet and J. F. Oliver,
{\it Europhys. Lett.} {\bf B77}, 51002 (2007).
%
\bibitem{Bai} Y. Bai {\it Phys.Lett.} {\bf B666}, 332 (2008).
%
\bibitem{Majumdar} D. Majumdar and A. Ghosal, {\it Mod. Phys. Lett.}
{\bf A23}, 2011 (2008).
%
\bibitem{DEramoF} F. D. Eramo {\it Phys. Rev.} {\bf D76}, 083522 (2007).
%
\bibitem{WanLeiGuo} W. L. Guo, X. Zhang, {\it Phys. Rev.} {\bf D79},
115023 (2009).
%
\bibitem{SilveiraV} V. Silveira and A. Zee, {\it Phys. Lett.} {\bf B161},
136 (1985).
%
\bibitem{Burgess} C. P. Burgess, M. Pospelov, T. ter Veldhuis, {\it Nucl. Phys.}
{\bf B619}, 709 (2001).
%
\bibitem{HolzE} D. E. Holz and A. Zee, {\it Phys. Lett.} {\bf B517}, 239
(2001).
%
\bibitem{McDonaldJ} J. McDonald,
{\it Phys. Rev.} {\bf D50},  3637 (1994).
%
\bibitem{PattB} B. Patt and F. Wilczek (2006), hep-ph/0605188.
%
\bibitem{BertolamiO} O. Bertolami, R. Rosenfeld, {\it Int. J. Mod. Phys.}
{\bf A23},  4817 (2008).
%
\bibitem{DavoudiaslG} H. Davoudiasl, R. Kitano, T. Li, and H. Murayama,
{\it Phys. Lett.} {\bf B609}, 117 (2005).
%
\bibitem{HeXG} X. G. He, T. Li, X. Q. Li,
and H.-C. Tsai, {\it Mod. Phys. Lett.} {\bf A22}, 117 (2005).
%
\bibitem{Eoi1} E. O. Iltan, {\it Eur. Phys. J.} {\bf C51}, 689 (2007).
%
\bibitem{BirdC} C. Bird, P. Jackson, R. Kowalewski and M. Pospelov,
{\it Phys. Rev. Lett.}  {\bf 93}, 201803 (2004).
%
\bibitem{BirdC2} C. Bird, R. Kowalewski and M. Pospelov,
{\it Mod. Phys. Lett.}  {\bf A21}, 457 (2006).
%
\bibitem{WAMP} D. N. Spergel et.al (WAMP Collaboration),
{\it Astrophys. J. Suppl. Ser.} {\bf 148} 175 (2003).
%
%
\bibitem{Gopalakrishna}  S. Gopalakrishna, S. Gopalakrishna, A. de Gouvea,
W. Porod, {\it JCAP}  {\bf 0605}, 005 (2006).
%
\bibitem{Gopalakrishna2}  S. Gopalakrishna, S. J. Lee, J. D.
Wells, hep-ph/0904.2007
%
\bibitem{KolbEW} E. W. Kolb and M. S. Turner, The Early Universe (Addison-
Wesley, Reading, MA, 1990).
%
\bibitem{Akerib}  D. S. Akerib  et.al. [CDMS Collaboration], {\it Phys. Rev.
Lett.}  {\bf 96}, 011302 (2006).
%
\bibitem{APierce}  A. Pierce, J. Thaler, {\it JHEP}  {\bf 0708}, 026 (2007).
%
\end{thebibliography}
\end{document}